\documentclass{JHEP3}

\usepackage{amssymb}
\usepackage{epsfig}

\title{General solution of scalar field cosmology with a (piecewise) exponential potential}
\author{Alexander A. Andrianov\\
V.A. Fock Department of Theoretical Physics, Saint Petersburg State University, 198904, S.Petersburg, Russia\\
Departament d'Estructura i Constituents de la Materia and Institut de Ci\`encies del Cosmos (ICCUB)
Universitat de Barcelona, 08028, Barcelona, Spain\\
E-Mail: \email{andrianov@bo.infn.it}}
\author{Francesco Cannata\\ INFN, sezione di Bologna, Via Irnerio 46, 40126 Bologna, Italy\\
E-Mail: \email{cannata@bo.infn.it}}
\author{Alexander Yu. Kamenshchik\\
Dipartimento di Fisica and INFN, Via Irnerio 46, 40126 Bologna,Italy\\
L.D. Landau Institute for Theoretical Physics of the Russian Academy of Sciences, Kosygin str.~2, 119334 Moscow, Russia\\ E-Mail: \email{kamenshchik@bo.infn.it}}

\abstract{We study in detail the general solution for a scalar field cosmology with an  exponential potential, correcting some imprecisions, encountered previously 
in the literature. In addition, we  generalize this solution for a piecewise exponential potential, which is continuous, but not smooth (with cusps).}


\preprint{}
\begin{document}

\section{Introduction}
Exact solutions of the Einstein equations play a very important role in cosmology, because they permit to study in a convenient way
the qualitative and quantitative features of the behavior of the universe as a whole. During the last decade the cosmological models
with scalar fields have acquired a great popularity. It is worth  mentioning various scenarios of the inflationary expansion of
the early universe \cite{inflation} or the the quintessence models of the dark energy \cite{dark} responsible for the
phenomenon of cosmic acceleration \cite{cosmic}. Nevertheless the number of known exact solutions for cosmological models
based on scalar fields is rather limited. One of such models is the flat Friedmann universe filled with a minimally coupled
scalar field with exponential potential. A particular solution for this model was known since the eighties and was studied in detail
\cite{power-law,power-law1,power-law2,power-law3,power-law4,power-law5}. This solution describes a power-law expansion of the universe. More recently, the general solution of the Einstein equations
for this model was constructed \cite{general}--\cite{general7}. This general solution was used for the description of such effects as transient acceleration,
and for the analysis of some models related to strings and branes.

Notwithstanding this activity, a good description of the general
solution, emphasizing its difference from the ``old'' particular solutions \cite{power-law,power-law1,power-law2,power-law3,power-law4,power-law5}, is not available to our knowledge.
In addition, all the considerations of the general solution for the scalar field with an exponential potential involved a real scalar fields
with real potentials. However, there are  attempts to consider cosmological models with  complex scalar fields and complex potentials.
Such models do not contradict to common sense, if the observable (first of all, geometrical) characteristics are real.
In our preceding papers \cite{we-PT},  inspired by the development of the PT-symmetric quantum theory \cite{PT,andr}, we have
developed models with complex potentials and have shown that they are rather convenient for the description of the so called
phantom cosmology \cite{phantom} , including such an enigmatic phenomenon as phantom divide line crossing \cite{divide}.

In the present paper we provide a comprehensible description of the general cosmological solution with an exponential potential;
then we explore the general solution for the phantom field. Finally, we describe the general exact cosmological solution for the
case of a piecewise exponential potential.

The paper is organized as follows: in Sec.~2 we describe
the general cosmological solution with an exponential potential in contrast to the ``old'' particular solution;
in Sec.~3 we present the phantom version of this solution;
in Sec. 4 we construct the general solution for a piecewise exponential potential with cusps;
the last section is devoted to Conclusions.

\section{The general solution with an exponential potential versus the ``old'' particular solution}
We shall study the flat Friedmann cosmological model described by the metric
\[
ds^2 = dt^2 - a^2(t)dl^2,
\label{Fried}
\]
where $a(t)$ is the cosmological radius of the universe.
The dynamics of the cosmological evolution is characterized by the Hubble variable
\[
h \equiv \frac{\dot{a}}{a},
\label{Hubble}
\]
which satisfies the Friedmann equation
\begin{equation}
h^2 = \varepsilon,
\label{Fried1}
\end{equation}
 where $\varepsilon$ is the energy density of the matter populating the universe and ``dot'' means the 
derivative with respect to the cosmic time parameter $t$.

We now basically follow the approach presented in Ref. \cite{general6}, adapting it to our purposes.
We shall consider the flat Friedmann universe (\ref{Fried}) filled with the minimally coupled scalar field
with the potential
\begin{equation}
V(\phi) = V_0 e^{\lambda \phi}.
\label{poten}
\end{equation}
Now the Friedmann equation (\ref{Fried1}) has the form
\begin{equation}
\frac{\dot{a}^2}{a^2} = \frac{\dot{\phi^2}}{2} + V_0 e^{\lambda\phi},
\label{Fried2}
\end{equation}
while the Klein-Gordon equation is
\begin{equation}
\ddot{\phi} + 3\frac{\dot{a}}{a}\dot{\phi} + \lambda V_0 e^{\lambda\phi} = 0.
\label{KG}
\end{equation}

\subsection{The particular exact solution}
A particular exact solution \cite{power-law,power-law1,power-law2,power-law3,power-law4,power-law5},
describing the power-law expansion of the universe can be found as follows:
Suppose that the scalar field has a time dependence
\begin{equation}
\phi(t) = \phi_0 \ln t + \phi_1,
\label{scal}
\end{equation}
while the cosmological radius behaves as
\begin{equation}
a = a_0 t^{k}.
\label{radius}
\end{equation}
In this case the Klein-Gordon (\ref{KG}) equation acquires the form
\begin{equation}
-\frac{\phi_0}{t^2} + \frac{3k\phi_0}{t^2} + \frac{\lambda V_0 e^{\lambda \phi_1}}{t^{-\lambda\phi_0}} = 0.
\label{KG1}
\end{equation}
It follows immediately that
\begin{equation}
\phi_0 = -\frac{2}{\lambda}.
\label{relation}
\end{equation}
Then
\begin{equation}
2(3k-1) = \lambda^2V_0e^{\lambda\phi_1}.
\label{relation1}
\end{equation}
The Friedmann equation (\ref{Fried2}) gives now
\begin{equation}
k^2 = \frac{2}{\lambda^2} + V_0e^{\lambda\phi_1}.
\label{relation2}
\end{equation}
Combining Eqs. (\ref{relation1}) and (\ref{relation2}) we obtain
\begin{equation}
k = \frac{6}{\lambda^2}.
\label{k}
\end{equation}
Substituting (\ref{k}) into the relation (\ref{relation2}) and requiring positivity of the coefficient
$V_0$, we obtain the following restriction on $\lambda$:
\begin{equation}
|\lambda| < 3\sqrt{2}.
\label{lambda}
\end{equation}
(Note that the limiting case $|\lambda| = 3\sqrt{2}$, would imply $k = 1/3$, i.e. the law of expansion of the universe
filled with stiff matter or massless scalar field, which in turn, means that $V_0 = 0$).
We can find also the value of the constant $\phi_1$:
\begin{equation}
\phi_1 = \frac{1}{\lambda}\ln \left(\frac{2(18-\lambda^2)}{\lambda^4 V_0}\right).
\label{phi1}
\end{equation}

Here, let us note that for the case of an imaginary $\lambda$ the  particular solution still exists and is
purely imaginary \cite{we-PT}. Here, both the potential and kinetic energy are real, but the latter
is negative and, hence, we encounter the phantom kind of matter.

It is convenient to write down again the explicit form of the exact particular solution:
\begin{equation}
\phi(t) = -\frac{2}{\lambda}\ln t + \frac{1}{\lambda}\ln \left(\frac{2(18-\lambda^2)}{\lambda^4 V_0}\right),
\label{phi-ex}
\end{equation}
\begin{equation}
h(t) = \frac{6}{\lambda^2 t}.
\label{Hubble-ex}
\end{equation}
The important feature of this solution is the rigid relation between the $\phi(t)$ and its time derivative
$\dot{\phi}(t)$:
\begin{equation}
\phi = -\frac{2}{\lambda}\ln \left(-\frac{2}{\lambda \dot{\phi}}\right) + \frac{1}{\lambda}\ln \left(\frac{2(18-\lambda^2)}{\lambda^4 V_0}\right).
\label{phi-phidot}
\end{equation}
 Thus, the particular exact solution corresponds to a particular choice of the initial conditions: at any moment of time
fixing the value of $\phi$, we automatically fix the value of $\dot{\phi}$.
It means that on the phase space of the problem under consideration the particular exact solution is described by
a unique trajectory. It is convenient to introduce a new variable:
\begin{equation}
\Phi \equiv \sqrt{V_0e^{\lambda\phi}}.
\label{Phi-def}
\end{equation}
Choosing the phase space variables as $\Phi$ and $\dot{\phi}$ we can calculate their relation:
\begin{equation}
\frac{\dot{\phi}}{\Phi} = -\frac{2\lambda}{\sqrt{36 - 2\lambda^2}} = const.
\label{phase-space}
\end{equation}
Thus, the trajectory on our phase space is a ray, beginning at the infinity and ending at the coordinate origin.
Correspondingly, the universe begins its evolution from a Big Bang singularity  and then undergoes an infinite expansion
(see Eq. (\ref{Hubble-ex}).

\subsection{The general solution}
Now, let us turn to the construction of the general exact solution.
It is convenient to introduce new variables, $u$ and $v$ such that
\begin{equation}
a^3 = e^{v+u},
\label{new1}
\end{equation}
\begin{equation}
\phi = A(v-u),
\label{new2}
\end{equation}
where $A$ is a coefficient to be defined.
Now the Friedmann equation (\ref{Fried2}) is
\begin{equation}
\frac{1}{9}(\dot{u}^2 +\dot{v}^2 + 2\dot{u}\dot{v}) = \frac{A^2}{2}(\dot{u}^2 +\dot{v}^2 - 2\dot{u}\dot{v}) + V_0 e^{\lambda A(v-u)}.
\label{Fried3}
\end{equation}
The Klein-Gordon equation (\ref{KG}) has now the form
\begin{equation}
A(\ddot{v}-\ddot{u}) + A(\dot{v}^2-\dot{u}^2) + \lambda V_0 e^{\lambda A(v-u)} = 0.
\label{KG3}
\end{equation}
Choosing the coefficient $A$ as
\begin{equation}
A = \frac{\sqrt{2}}{3},
\label{A}
\end{equation}
we give a ``light-cone form'' to Eq. (\ref{Fried3}):
\begin{equation}
\dot{v}\dot{u} = \frac94 V_0 e^{\frac{\sqrt{2}}{3}\lambda (v-u)}.
\label{Fried4}
\end{equation}
Now we want to simplify further this equation, choosing a new time parameter $\tau$.
Rewriting Eq. (\ref{Fried4}) as
\begin{equation}
v' u' \dot{\tau}^2 = \frac94 V_0 e^{\frac{\sqrt{2}}{3}\lambda (v-u)}
\label{Fried5}
\end{equation}
where prime denotes the derivative with respect to $\tau$.
It is convenient to choose this new time parameter such that
\begin{equation}
\dot{\tau}= \frac32 \sqrt{V_0} e^{\frac{\lambda\phi}{2}} = \frac32 \sqrt{V_0} e^{\frac{\lambda(v-u)}{\sqrt{6}}},
\label{tau}
\end{equation}
we come to
\begin{equation}
v' u' = 1.
\label{Fried6}
\end{equation}
Correspondingly, the Klein-Gordon equation (\ref{KG3}) is now
\begin{equation}
v'' - u'' + \frac{\sqrt{2}\lambda}{6}(v'-u')^2 + (v'^2-u'^2) + \frac{2\sqrt{2}}{3}\lambda = 0.
\label{KG4}
\end{equation}
Substituting $u' = 1/v'$ from Eq. (\ref{Fried6}) into Eq. (\ref{KG4}) one arrives to
\begin{equation}
v'' + \left(1 + \frac{\sqrt{2}\lambda}{6}\right) v'^2 + \left(\frac{\sqrt{2}\lambda}{6} - 1\right) = 0.
\label{KG5}
\end{equation}
Introducing a new variable
\begin{equation}
x \equiv  v'
\label{x}
\end{equation}
we rewrite the Klein-Gordon equation (\ref{KG5}) in the Riccati form:
\begin{equation}
x' + \left(1 + \frac{\sqrt{2}\lambda}{6}\right) x^2 + \left(\frac{\sqrt{2}\lambda}{6} - 1\right) = 0.
\label{Ricc}
\end{equation}
Making the substitution
\begin{equation}
x = \frac{1}{\left(1 + \frac{\sqrt{2}\lambda}{6}\right)}\frac{y'}{y},
\label{Ricc1}
\end{equation}
we obtain the second-order linear differential equation
\begin{equation}
y'' + \left(\frac{\lambda^2}{18}-1\right) y = 0.
\label{Ricc2}
\end{equation}
We should consider separately two cases : the ``hyperbolic'' one, when the constant $\lambda$ satisfies the
condition (\ref{lambda}) and ``trigonometric'' one, when the condition (\ref{lambda}) is not satisfied.

\subsubsection{The hyperbolic case}
The solution of Eq. (\ref{Ricc2}) in this case is
\begin{equation}
y(\tau) = B e^{\kappa \tau} + C e^{-\kappa \tau},
\label{Ricc3}
\end{equation}
where
\begin{equation}
\kappa \equiv \sqrt{1-\frac{\lambda^2}{18}}.
\label{kappa}
\end{equation}
Then, substituting (\ref{Ricc3}) into Eqs. (\ref{x}) and (\ref{Ricc1}) after an elementary integration we find
\begin{equation}
v = \frac{1}{\left(1 + \frac{\sqrt{2}\lambda}{6}\right)}\ln (B e^{\kappa \tau} + C e^{-\kappa \tau}) + v_0.
\label{v-sol}
\end{equation}
Using the relation (\ref{Fried6}) we can analogously find
\begin{equation}
u = \frac{1}{\left(1 - \frac{\sqrt{2}\lambda}{6}\right)}\ln (B e^{\kappa \tau} - C e^{-\kappa \tau}) + u_0.
\label{u-sol}
\end{equation}

Now, remembering the formulae (\ref{new2}) and (\ref{A}), we can write
\begin{eqnarray}
&&\phi = \frac{\sqrt{2}}{3}\left((v_0-u_0)+\frac{1}{\left(1 + \frac{\sqrt{2}\lambda}{6}\right)}\ln (B e^{\kappa \tau} + C e^{-\kappa \tau})
\right.\nonumber \\
&&\left.-\frac{1}{\left(1 - \frac{\sqrt{2}\lambda}{6}\right)}\ln (B e^{\kappa \tau} - C e^{-\kappa \tau})\right).
\label{phi3}
\end{eqnarray}

We have to consider separately two cases: $C = 0$ and $C \neq 0$. If $C = 0$ the solution (\ref{phi3}) can be rewritten as
\begin{equation}
\phi = \phi_2 - \frac{2\lambda\tau}{9\kappa},
\label{phi4}
\end{equation}
where $\phi_2$ is an arbitrary constant. Substituting Eq.(\ref{phi4}) into Eq. (\ref{tau}), connecting the time parameter
$\tau$ with the cosmic time $t$ we obtain the following equation
\begin{equation}
\frac{d \tau}{d t} = \frac32\sqrt{V_0}\exp\left(\frac{\lambda\phi_2}{2}-\frac{\lambda^2\tau}{9\kappa}\right).
\label{tau1}
\end{equation}
Integrating the equation (\ref{tau1}) we find the parameter $\tau$ as a function of $t$ and substituting it into
Eq. (\ref{phi4}) we come to the particular exact solution (\ref{phi-ex}). The dependence on the constant $\phi_2$ disappears.
Thus, the case $C = 0$ coincides with the particular exact solution described in the subsection 2.1.
One can easily show that the case $B=0$
corresponds to the particular solution, described in the preceding subsection, but it refers to a universe contracting
from a state with an infinite radius to the Big Crunch singularity.

In the case when $C\neq 0, B \neq 0$ the solution (\ref{phi3}) can be rewritten as
\begin{eqnarray}
&&\phi = \phi_3 - \frac{2\lambda\tau}{9\kappa} +\frac{\sqrt{2}}{3\left(1 + \frac{\sqrt{2}\lambda}{6}\right)}\ln \left(1 + \frac{C}{B} e^{-2\kappa \tau}\right)\nonumber \\
&&-\frac{\sqrt{2}}{3\left(1 - \frac{\sqrt{2}\lambda}{6}\right)}\ln \left(1- \frac{C}{B} e^{-2\kappa \tau}\right),
\label{phi5}
\end{eqnarray}
where $\phi_3$ is a constant.
As a matter of fact, here  there are two families of solutions.
We can rescale the factor $\frac{C}{B}$, that is equivalent to a shift of the variable $\tau$.
Let us consider first the case $C/B = 1$. Then we come
to a more simple expression:
\begin{eqnarray}
&&\phi = \phi_3 - \frac{2\lambda\tau}{9\kappa} +\frac{\sqrt{2}}{3\left(1 + \frac{\sqrt{2}\lambda}{6}\right)}\ln (1 + e^{-2\kappa \tau})
-\frac{\sqrt{2}}{3\left(1 - \frac{\sqrt{2}\lambda}{6}\right)}\ln (1- e^{-2\kappa \tau}).
\label{phi6}
\end{eqnarray}

While the particular solution (\ref{phi-ex}) describes the unique trajectory in the two-dimensional phase space
of variables $(\Phi,\dot{\phi})$ which can be represented by Eq. (\ref{phi-phidot}) or by Eq. (\ref{phase-space})
the general solution (\ref{phi6}) describes a family of solutions, parameterized by the constant $\phi_3$. Indeed,
in this case a rigid relation between $\phi$ and $\dot{\phi}$ is absent. If one fixes the value of $\phi$ then
the corresponding value of $\dot{\phi}$ is given by $\dot{\phi}=\phi'\cdot\dot{\tau}$. As it follows from Eq. (\ref{tau})
 fixing  the value of $\phi$ determines the value of $\dot{\tau}$. However, changing the value of the constant $\phi_3$
while preserving the value of $\phi$ is equivalent to shifting the value of $\tau$, and, hence, changing the value of
$\phi'$ and $\dot{\phi}$.  One can represent it in a different manner. Let us note that the value of $\tau$ in Eq. (\ref{phi6})
is changed in the interval $0 \leq \tau < \infty$. We can now calculate the relation between $\dot{\phi}$ and the phase space
variable $\Phi$ analogous to (\ref{phase-space}). One obtains
\begin{equation}
\frac{\dot{\phi}}{\Phi} = -\frac{\lambda}{3\kappa} - \frac{\sqrt{2}e^{-2\kappa\tau}}{\kappa(1-e^{-4\kappa\tau})}\left(2+\frac{\sqrt{2}\lambda}{3}e^{-2\kappa\tau}\right).
\label{phase-space1}
\end{equation}
Note, that the value of both $\phi'$ and $\dot{\phi}$ is negative during the cosmological evolution.

In contrast to the formula (\ref{phase-space}) the formula (\ref{phase-space1}) does not describe a ray. The corresponding
curve begins at the point $\Phi(\tau=0) = +\infty, \dot{\phi}(\tau=0) = -\infty$ and ends at $\tau = +\infty$ at the origin
of the coordinates. It is easy to see from Eq. (\ref{phase-space1}) that the value of $\frac{\dot{\phi}}{\Phi}$ at the moment
$\tau = 0$ tends to $-\infty$. That means that at the beginning of the cosmological evolution the kinetic term dominates the
potential term. It is well-known that in this case (the case of the massless scalar field or, equivalently, the stiff matter)
the Hubble parameter behaves as
\begin{equation}
h(t) \rightarrow \frac{1}{3t},\ \ t \rightarrow 0.
\label{Hubble-BB}
\end{equation}
The direct evaluation obtained by the substitution of the exact solution
(\ref{phi6}) into the Friedmann equation (\ref{Fried2}) and using the relation (\ref{tau}) in the limiting case $t \rightarrow 0$ confirms the behaviour (\ref{Hubble-BB}).
At the end of the evolution $\tau = +\infty$, the relation (\ref{phase-space1})  coincide with that of the ray (\ref{phase-space}). Thus, the curves, corresponding to the general solution (\ref{phi6}) (excluding the particular solution (\ref{phi-ex})) begins at the Big Bang singularity, where the kinetic term dominates the potential one, which in the phase plane $(\Phi,\dot{\phi})$ corresponds to the asymptote coinciding with the negative semi-axis $\Phi = 0$.  Then, the curves
(\ref{phase-space1}) are located under the ray (\ref{phase-space}) and conclude their evolution at the coordinate origin,
which corresponds to an eternal expansion.

Now we can consider the second family of solutions, which can be obtained by putting  $C/B = -1$. The time parameter again runs between $0$ and $+\infty$. However, the expression
for the scalar field is now
\begin{eqnarray}
&&\phi = \phi_3 - \frac{2\lambda\tau}{9\kappa} +\frac{\sqrt{2}}{3\left(1 + \frac{\sqrt{2}\lambda}{6}\right)}\ln (1 - e^{-2\kappa \tau})
-\frac{\sqrt{2}}{3\left(1 - \frac{\sqrt{2}\lambda}{6}\right)}\ln (1+ e^{-2\kappa \tau}).
\label{phi60}
\end{eqnarray}
This solution also describes an evolution from the Big Bang to an infinite expansion, however, now at the
beginning of the evolution the scalar field $\phi \rightarrow -\infty$ and the potential energy is equal
to zero. During the evolution, the scalar field is growing, then at some moment its time derivative changes the sign
and it becomes decreasing. At the end of the evolution (an infinite expansion) when $\tau \rightarrow \infty$ the scalar
field again tends to $-\infty$ and both the kinetic and potential energy vanish.

The particular exact solution plays a role of the separatrix between these two families of solutions. One can ask how to
understand to which exactly family belong a certain trajectory being given the values of the scalar field $\phi$ and its
time derivative (with respect to the cosmic time $t$) at some moment of time. The answer is simple. It is enough to
calculate the derivative of the scalar field with respect to the parametric time $\tau$, which is
$\phi' = \frac{2\dot{\phi}}{3\sqrt{V(\phi)}}$. If its value is less than $-\frac{2\lambda}{9\kappa}$. then the trajectory
belongs to the first family. In the opposite case it belongs to the second one.
In Figure 1, we sketch the trajectory describing the particular exact solution and two trajectories, representing two families of the
general solution for the hyperbolic case (for definiteness, we have chosen for all figures the subcase $\lambda > 0$).

\begin{figure}[h]
\centerline{\epsfxsize 7.5cm \epsfbox{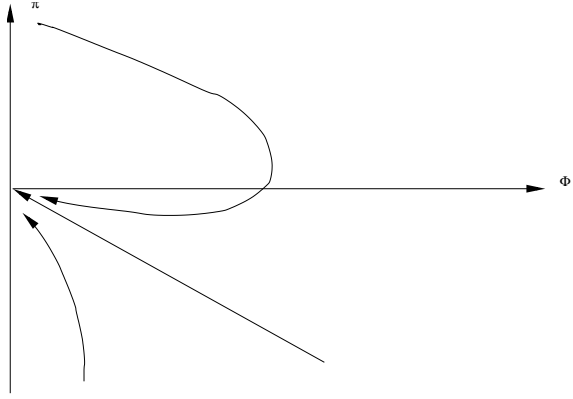}}
\caption{\small The phase space diagram $\Phi,\pi$, where $\pi$ stays for the velocity $\pi = \dot{\phi}$. The straight line represents the exact particular solution for the hyperbolic case, while two
curved lines represents solutions, belonging to two families of general hyperbolic solutions described in this subsection.}
\end{figure}

\subsubsection{The trigonometric case}
Now we consider the case when $\lambda^2 > 18$. In this case
the solution of Eq. (\ref{Ricc2}) can be chosen as $\sin\omega\tau$ or $\cos\omega\tau$.
However, in contrast with the hyperbolic case, these two choices do not imply the real difference between
cosmological evolutions, because the transition from one to another corresponds to the shift of the parametric time
interval. Thus, we shall choose
\begin{equation}
y(\tau) = D \sin\omega \tau,
\label{Ricc4}
\end{equation}
where
\begin{equation}
\omega \equiv \sqrt{\frac{\lambda^2}{18}-1},
\label{omega}
\end{equation}
and $D$ and $\alpha_0$ are real constants.
Then using the definitions of (\ref{Ricc1}) and (\ref{x}) we get
\begin{equation}
v = \frac{1}{1 + \frac{\sqrt{2}\lambda}{6}} \ln \sin\omega \tau + v_0,
\label{v-trig}
\end{equation}
and similarly
\begin{equation}
u = -\frac{1}{\frac{\sqrt{2}\lambda}{6}-1}\ln \cos\omega \tau + u_0.
\label{u-trig}
\end{equation}
Finally, the general solution for this trigonometric case looks like
\begin{equation}
\phi = \phi_4 + \frac{\sqrt{2}}{3}\left(\frac{1}{1 + \frac{\sqrt{2}\lambda}{6}} \ln \sin\omega \tau+
\frac{1}{\frac{\sqrt{2}\lambda}{6}-1}\ln \cos\omega \tau\right),
\label{solution-trig}
\end{equation}
where $\phi_4$ is a constant. We note that the expression describes an expanding universe
if  the parametric time $\tau$ is changing inside the interval
\begin{equation}
0 \leq \omega\tau \leq \frac{\pi}{2}.
\label{time-trig}
\end{equation}
The relation between the parametric time $\tau$ and the cosmic time $t$ is
\begin{equation}
\dot{\tau} = \frac32\sqrt{V_0}e^{\lambda\phi_4/2}(\sin \omega\tau)^{\frac{\frac{\sqrt{2}\lambda}{6}}{1 + \frac{\sqrt{2}\lambda}{6}}}(\cos \omega\tau)^{\frac{\frac{\sqrt{2}\lambda}{6}}{\frac{\sqrt{2}\lambda}{6}-1}}.
\label{tau-trig}
\end{equation}
We shall present  also
\begin{equation}
\phi' = \frac{2\sqrt{2}\left(\frac{\sqrt{2}\lambda}{6}\cos2\omega\tau-1\right)}{3\omega \sin 2\omega\tau}.
\label{phi-prime-trig}
\end{equation}
Using the formulae (\ref{solution-trig}), (\ref{tau-trig}) and (\ref{phi-prime-trig}) one can find
that when $\tau$ runs from $0$ to $\pi/2\omega$, the cosmic time runs from $0$ to $\infty$.
At the beginning of the evolution the velocity $\dot{\phi}$ is infinite, while the potential energy is zero.
The direct calculation shows that the Hubble parameter behaves at the beginning of the evolution as
$h = \frac{1}{3t}$ as it should be for universe, filled with a massless scalar field arising
from the Big Bang singularity. At the end of the evolution the potential
energy again tends to zero, while the velocity $\dot{\phi}$ being negative also tends to zero.
(Note that the time derivative with respect to the parametric time $\tau$ diverges when
$\omega\tau \rightarrow \pi/2$ and one can think that here we again encounter the singularity as was mentioned in
\cite{general6}. However, it is not the case, because the expression $\dot{\tau}$ tends to zero more rapidly and thus,
the kinetic energy of the scalar field tends to zero).
This situaiton corresponds to
an eternal expansion. From Eq. (\ref{phi-prime-trig}) it follows that the velocity changes sign at the moment
\begin{equation}
\tau_1 = \frac{1}{2\omega}{\rm arccos}\frac{3\sqrt{2}}{\lambda}.
\label{change}
\end{equation}
In Figure 2 we represent a typical trajectory for the trigonometric case.

\begin{figure}[h]
\centerline{\epsfxsize 7.5cm \epsfbox{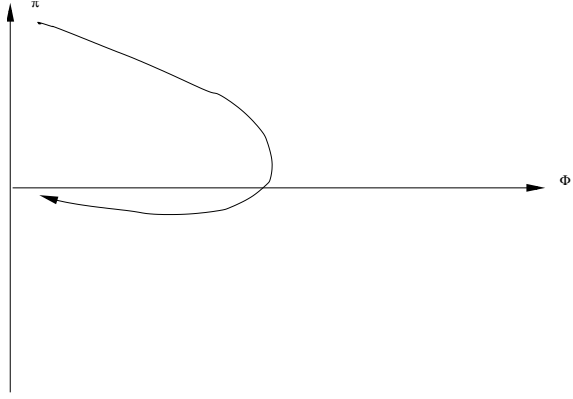}}
\caption{\small The phase space diagram $\Phi,\pi$, where $\pi$ stays for the velocity $\pi = \dot{\phi}$. The curved line represents one of trajectories, belonging to the family,  given by the general solution for the trigonometric case.}
\end{figure}

\subsubsection{The case $\lambda = \sqrt{18}$}
In this case the equation for the variable $v$ is the following :
\begin{equation}
v'' + 2v'^2 = 0,
\label{limit}
\end{equation}
whose solution is
\begin{equation}
v = \frac12\ln \tau + v_0.
\label{v-lim}
\end{equation}
Correspondingly
\begin{equation}
u = \tau^2 + u_0
\label{u-lim}
\end{equation}
and
\begin{equation}
\phi = \phi_5 +\frac{\sqrt{2}}{3}\left(\frac12\ln \tau - \tau^2\right).
\label{phi-lim}
\end{equation}
The relation between $\tau$ and $t$ is now
\begin{equation}
\dot{\tau} = \frac32\sqrt{V_0}e^{3\sqrt{2}\phi_5/2}\sqrt{\tau} e^{-\tau^2}.
\label{time-lim}
\end{equation}
It is easy to see that $\tau$ runs from $0$ to $\infty$ and the cosmic time $t$ has the same range.
Then
\begin{equation}
\phi' = \frac{\sqrt{2}}{6\tau} - \frac{2\sqrt{2}\tau}{3}.
\label{phi-prime-lim}
\end{equation}
It is easy to see that at the beginning and at the end of the evolution the field $\phi$ tends to $-\infty$ and hence the
potential energy vanishes. Then, at the beginning of the evolution the $\dot{\phi}$ is infinite and that means that
the universe is born from the Big Bang singularity, while at the end of the evolution the negative time derivative of the
scalar field $\dot{\phi}$ vanishes and the universe is expanding infinitely.

It is curious to note that the particular solution described in the subsection 2.1 exists in the case $\lambda^2 = 18$.
In this case, as it follows from Eqs. (\ref{relation1})--(\ref{k}),
\begin{equation}
V_0 = 0
\label{V0}
\end{equation}
and this is simply the case of the massless scalar field. This particular case cannot be extracted from the general solution
(\ref{phi-lim}) because the very existence of this general solution is based on the introduction of the parametric time
$\tau$ which is defined by means of Eq. (\ref{tau}) which is senseless when $V_0 = 0$. Thus, the particular solution
for the hyperbolic case is contained in the general formula (\ref{phi3}). In the case $\lambda^2 = 18$ the particular solution
exists, but it should be treated separately, while the trigonometric case does not have a particular solution.

\section{The  exact solution for a phantom scalar field with an exponential potential}
Considering the phantom scalar field with a negative kinetic term, we shall have the following Friedmann
\begin{equation}
\frac{\dot{a}^2}{a^2} = -\frac{\dot{\phi^2}}{2} + V_0 e^{\lambda\phi},
\label{Fried-phan}
\end{equation}
and Klein-Gordon
\begin{equation}
\ddot{\phi} + 3\frac{\dot{a}}{a}\dot{\phi} - \lambda V_0 e^{\lambda\phi} = 0
\label{KG-phan}
\end{equation}
equations. We shall first present the particular solution for this case.
\subsection{The particular exact solution for the phantom case}
As in the subsection 2.1, we shall look for the solution for the phantom scalar field, which depends logarithmically
on the cosmic time $t$, while the scale factor undergoes a power-law expansion (or contraction). If we would like to
consider an expanding universe, then the solution will be
\begin{equation}
\phi(t) = -\frac{2}{\lambda}\ln (-t) + \frac{1}{\lambda}\ln \left(\frac{2(18+\lambda^2)}{\lambda^4 V_0}\right),
\label{part-phan}
\end{equation}
 where $t$ is running from $-\infty$ to $0$.
The Hubble parameter is now
\begin{equation}
h(t) = - \frac{6}{\lambda^2 t}.
\label{Hubble-phan}
\end{equation}
Thus, the formulae (\ref{part-phan}) and (\ref{Hubble-phan}) describe a cosmological evolution which
begins at $t = -\infty$ with infinitely small cosmological radius and ends at $t = 0$, encountering a Big Rip singularity.
However, another particular solution for the phantom case does exist. Here the formula for the Hubble factor
 is the same (\ref{Hubble-phan}), but in the formula for the scalar field (\ref{part-phan}) $\ln t$  instead of $\ln (-t)$
enters. This solution describes the contraction of the universe, which begins at the moment $t =0$ in the ``anti-Big Rip'' singularity characterized by an infinite radius and infinite negative Hubble parameter, and ends at $t = +\infty$
with an endless contraction.

\subsection{The general exact solution for the phantom case}
As in the subsection 2.2 we shall introduce the variables $v$ and $u$ (see Eqs. (\ref{new1}),(\ref{new2}), (\ref{A})).
However, because of the negative sign of the kinetic term in the right-hand side of Eq. (\ref{Fried-phan}) we shall
obtain instead of (\ref{Fried4})
\begin{equation}
\dot{v}^2 + \dot{u}^2 = \frac92V_0 e^{\lambda\phi}.
\label{Fried-phan1}
\end{equation}
It is convenient now to introduce a complex variable
\begin{equation}
z \equiv  \frac{1}{\sqrt{2}}(v + iu).
\label{z-def}
\end{equation}
 Now, Eq. (\ref{Fried-phan1}) looks like
\begin{equation}
\dot{z}\dot{\bar{z}} = \frac94V_0e^{\lambda\phi},
\label{z-eq}
\end{equation}
where ``bar'' stands for the complex conjugation.
Introducing now the time parameter $\tau$ like in Eq. (\ref{tau}) we obtain
\begin{equation}
z'\bar{z}' = 1.
\label{z-eq1}
\end{equation}
Rewriting the Klein-Gordon equation (\ref{KG-phan}) and taking into account the relation (\ref{z-eq1}) we
come to
\begin{equation}
z'' +\frac{\sqrt{2}(1-i)}{2}\left[z'^2\left(1+\frac{\sqrt{2}\lambda i}{6}\right) - i -\frac{\sqrt{2}\lambda}{6}\right] = 0.
\label{KG-z}
\end{equation}
Introducing the function $f$ such that
\begin{equation}
z' \equiv \frac{1}{\alpha}\frac{f'}{f},
\label{f-phan}
\end{equation}
 where
\begin{equation}
\alpha \equiv \frac{\sqrt{2}(1-i)}{2}\left(1+\frac{\sqrt{2}\lambda i}{6}\right).
\label{alpha}
\end{equation}
The auxiliary function $f$ satisfies the following equation
\begin{equation}
f'' - \tilde{\kappa}^2f = 0,
\label{f-phan1}
\end{equation}
where
\begin{equation}
\tilde{\kappa} \equiv \sqrt{1+\frac{\lambda^2}{18}}.
\label{kappa-phan}
\end{equation}
The general solution of Eq. (\ref{f-phan1}) is
\begin{equation}
f(\tau) = Fe^{\tilde{\kappa}\tau} + Ge^{-\tilde{\kappa}\tau}.
\label{gen-phan-f}
\end{equation}
Hence,
\begin{equation}
z = \frac{1}{\alpha}\ln f.
\label{z-f}
\end{equation}
Now, we can find the expression for the scalar field
\begin{equation}
\phi(\tau) = \frac{2\lambda}{9\tilde{\kappa}^2}\ln|f| -\frac{2\sqrt{2}}{3\tilde{\kappa}^2}\arg f + const.
\label{phi-phan4}
\end{equation}
Now, substituting the expression for $f$ (\ref{gen-phan-f}) into the expression for $z$ (\ref{z-f}) and substituting the latter
into the condition (\ref{z-eq1}) we come to the consistency equation
\begin{equation}
F\bar{G} + G\bar{F} = 0.
\label{consist}
\end{equation}
These equation can be satisfied if one of the coefficients is equal to zero or if the difference of their phases is equal to
$\pi/2$.

First, we consider  the case $G = 0$.
Then
\begin{equation}
\phi(\tau) = \frac{2\lambda\tau}{9\tilde{\kappa}} + \phi_8.
\label{phi-phan6}
\end{equation}
Substituting this expression into Eq. (\ref{tau}) we can find $\tau$ as a function of the cosmic time $t$.
\begin{equation}
\tau = -\frac{9\tilde{\kappa}}{\lambda^2}\left(\ln(-t) + \frac{\lambda\phi_9}{2} + \ln\left(\frac{\sqrt{V_0}\lambda^2}{6\tilde{\kappa}}\right)\right).
\label{tau-t-phan}
\end{equation}
Substitung this expression into the solution (\ref{phi-phan6}) we come to the particular exact solution
(\ref{part-phan}) as it should be and in complete analogy with the hyperbolic case. It is easy to see from
Eq. (\ref{tau-t-phan}) that when the cosmic time runs from $-\infty$ to $0$, the parameter $\tau$ runs from
$-\infty$ to $+\infty$, while the cosmic time $t$ runs from $t=-\infty$ to $t =0$. In contrast to the hyperbolic case
considered in subsubsection 2.2.1 (cf. Eq. (\ref{phi3})) we can here put also $F =0$, while $G \neq 0$. In this case we reproduce
the second particular solution, describing an infinite contraction of the universe, which begins at the anti-Big Rip singularity.

Now, we consider the case when both the constants $F$ and $G$ are different from zero. We can choose one of these constants,
say $F$ equal to 1 while $G = i$.

\begin{equation}
\phi(\tau) = \phi_{10} + \frac{2\lambda}{9\tilde{\kappa}^2}\ln \cosh 2\tilde{\kappa}\tau -
\frac{2\sqrt{2}}{3\tilde{\kappa}^2}{\rm arctan}(e^{-2\tilde{\kappa}\tau}).
\label{phi-phan7}
\end{equation}
 Here, the time parameter $\tau$ runs from $-\infty$ to $+\infty$. When $\tau \rightarrow -\infty$ the field $\phi \rightarrow +\infty$, while when $\tau \rightarrow +\infty$ the field $\phi \rightarrow +\infty$ again. Thus, the beginning and the end
of the cosmological evolution are characterized by a positive infinite value of the scalar field, and, hence by the positive
infinite value of the potential $V$. It is useful to write down also the explicit expression for the Hubble parameter:
\begin{equation}
h(\tau) = \frac{2\dot{\tau}}{3\tilde{k}}\left(\frac{\sinh 2\tilde{\kappa}\tau-\frac{\sqrt{2}\lambda}{6}}{\cosh 2\tilde{\kappa}\tau}\right).
\label{Hubble-phan3}
\end{equation}
It is easy to see that at $\tau \rightarrow -\infty$ the Hubble parameter $h \rightarrow -\infty$ at $\tau \rightarrow +\infty$
the Hubble parameter $h \rightarrow +\infty$ and at $\tau = \frac{1}{2\tilde{\kappa}}{\rm arcsinh}\frac{\sqrt{2}\lambda}{6}$ the Hubble parameter changes the sign and, hence, the universe
passes through the point of minimal contraction. The complete cosmological evolution involves a finite period of the cosmic time $t$.
 One can say that these finite time
evolutions represent the phantom counterpart of the well-known cosmological evolutions, which begin in the Big Bang singularity,
reach the point of maximal expansion and then have a stage of contraction culminating in the Big Crunch singularity.
These evolutions can be observed not only in the closed Friedmann models, but also in the flat Friedmann models with
the standard scalar field with a negative potential. (As far as we could understand this solution beginning from
``anti-Big-Rip'' singularity and ending in the Big Rip singularity, passing through the point of minimal contraction
was not considered in paper \cite{Elizalde} devoted to the study of phantom solutions with exponential potentials).

In Figure 3 we represent two particular exact solution and a typical example of general solution for the phantom case.

\begin{figure}[h]
\centerline{\epsfxsize 7.5cm \epsfbox{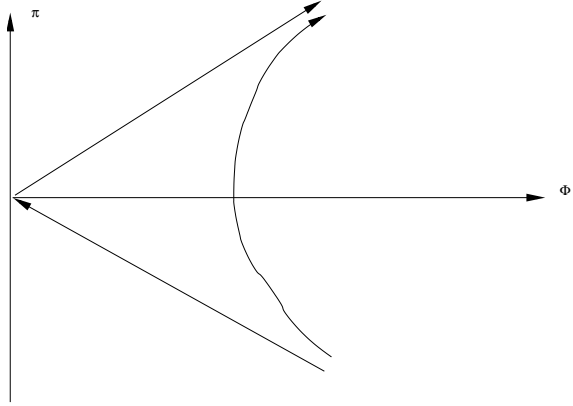}}
\caption{\small The phase space diagram $\Phi,\pi$, where $\pi$ stays for the velocity $\pi = \dot{\phi}$. Two straight line trajectories describe
two particular solutions, corresponding to the expanding and contracting universes. The curved line represents a trajectory, belonging to the family of those evolving from the anti-Big Rip singularity to the Big Rip singularity, passing through the point of minimal contraction of the universe.}
\end{figure}

\subsection{The complexification of the scalar field and of its potential}
In principle, the phantom solutions can be obtained from the hyperbolic non-phantom solutions by some kind of analytic continuation.
Let us consider the case when the parameter $\lambda = i\Lambda$ (where $\Lambda$ is real)  is imaginary. One can see that this case is a ramification of the hyperbolic case, because the parameter $\kappa=\sqrt{18+\Lambda^2}$ (\ref{kappa}) is well defined. However, now also the solution of Klein-Gordon
equation should be purely imaginary. Thus, taking Eq. (\ref{Ricc2}) with $\kappa=\sqrt{18+\Lambda^2}$ we obtain the general
equation of the hyperbolic type, which can be written as was don in Eq. (\ref{phi3}) with arbitrary coefficients $B$ and $C$.
Now, choosing these coefficients as $B =1$ and $C = i$, we come to purely imaginary solution, which behaves
just like our phantom solution (\ref{phi-phan7}). multiplied by $i$. Substituting this solution into the Friedmann equation
we obtain an equation of the evolution of the universe, filled with a phantom scalar field. Naturally, all the observables
appear to be real, and when necessary also positive (cf. \cite{we-PT}).

\section{General exact solutions for piecewise exponential potentials}
In this section we analyze the generalization of the previous results for the case of the potentials, represented by
piecewise functions, where all the pieces are exponential. We limit ourselves by continuous (but not smooth) potentials.
Such potentials can be written down as
\begin{eqnarray}
&&V^{(N)}(\phi) = \theta(\phi_1-\phi)V_0e^{\lambda_0\phi}\nonumber \\
&&+\theta(\phi-\phi_1)\theta(\phi_2-\phi)V_1e^{\lambda_1\phi}+\cdots\nonumber \\
&&+\theta(\phi-\phi_{N-1})\theta(\phi_N-\phi)V_{N-1}e^{\lambda_{N-1}\phi}\nonumber\\
&&+\theta(\phi-\phi_N)V_Ne^{\lambda_N\phi}.
\label{piecewise}
\end{eqnarray}
 where $\theta$ is  the Heaviside function
and where
\begin{equation}
-\infty <\phi_1 < \phi_2 < \cdots < \phi_N< \infty,
\end{equation}
and
\begin{equation}
V_ke^{\lambda_k\phi_{k+1}}=V_{k+1}e^{\lambda_{k+1}\phi_{k+1}},\ 0 \leq k \leq N-1,
\label{piecewise1}
\end{equation}
The general exact solutions exist also in this case and can be realized by means of the matching across the values
of scalar field, where the exponent $\lambda$ changes its value. This matching involves not only the value of the
scalar field, but also the continuity of $\dot{\phi}$ and, hence, $\phi'$. In the opposite case, one would
encounter a jump in the value of the Hubble parameter, following from the Friedmann equation, which seems
unphysical.
The simplest potential with cusp of this type is $V(\phi) = V_0e^{\lambda|\phi|}$, which can be obtained from our general
potential (\ref{piecewise}) by choosing $N = 1, \phi_1 = 0, V_1 = V_0$ and $\lambda_0 = -\lambda_1 = -\lambda$.

To get an idea of new possibilities let us consider, for example, the matching across the value $\phi = \phi_1$ and focus on an expanding universe.
If $\lambda_0$ satisfies the relation
$\lambda_0^2 > 18$ (the trigonometric case) two situations are possible: the derivative $\phi'$ can be positive
(an initial part of the evolution) and it can be negative (the final part of the evolution) as follows from Eq.
(\ref{phi-prime-trig}).
If the derivative is negative, that means that we cross the value $\phi=\phi_1$ from the right ( from the side of larger
values of $\phi$) and then the value of $\phi$ is decreasing indefinitely.

If this derivative is positive the field crosses the point $\phi = \phi_1$ and enters
the range of values between $\phi_1$ and
$\phi_2$, where the exponent $\lambda$ is equal to $\lambda_1$.  If $\lambda_1$ is also trigonometric, then
the universe continues its expanding evolution, with the scalar field given by Eq. (\ref{solution-trig}) with $\lambda =\lambda_1$
and the corresponding initial conditions. Then, again two situations are possible - the field $\phi$ continues growing arriving
to the value $\phi = \phi_2$, where another change of regime occurs, or it can arrive at some moment to the value
$\phi = \phi_* < \phi_2$ while $\phi' = 0$. After that the field begins decreasing arriving at the value $\phi = \phi_1$ with
a negative derivative $\phi'$ and enters into the range of values with $\phi < \phi_1$, decreasing indefinitely until $-\infty$,
which corresponds to an eternal expansion with the vanishing value of the Hubble parameter.

If the value of $\lambda_1^2 < 18$ (hyperbolic case) and if the value of $\phi'$ is less than $-\frac{2\lambda_1}{9\kappa_1}$ then the solution in the range $\phi_1 \leq \phi \leq \phi_2$ is the hyperbolic solution of the first kind (described in the
subsubsection 2.2.1) and the field enters into the region $\phi < \phi_1$ from the right, continuing to decrease indefinitely.
If the value of $\phi'$ at the point, where $\phi = \phi_1$ is negative but  larger than $-\frac{2\lambda_1}{9\kappa_1}$,
then the solution in the region $\phi_1 \leq \phi \leq \phi_2$ is the hyperbolic solution of the second kind and again
the scalar field after the crossing the value $\phi = \phi_1$ decreases indefinitely. If instead the value of $\phi'$ at the
point of transition is positive the fields enters into the region $\phi_1 \leq \phi \leq \phi_2$ increasing and following the
hyperbolic solution of the second kind. Then, again, two situations are possible: the field $\phi$ growing can achieve the value $\phi_2$ with the subsequent change of the regime or it can begin decreasing after achieving some maximal value $\phi_{max} < \phi_2$, entering back into the region $\phi < \phi_1$.

It is important to pay a special attention to the situation when the field $\phi$ arrives at the value $\phi = \phi_1$ with
$\phi' = 0$. As follows from Eq. (\ref{phi-prime-trig}) that means that the field $\phi$ has reached its maximal value and
begin decreasing. Thus, in this case the change of regime does not occur.
In our preceding papers \cite{we,we1} (see also \cite{Yurov}), in was shown that the phantom divide  line crossing effect, provoked by self-conversion of a non-phantom scalar field into a phantom scalar field (or vice versa) can take
place in the models with a unique minimally coupled scalar field. To realize this effect it is necessary to consider the cusped potentials
and some particular initial conditions. In other words, it is necessary to approach the cusp with a vanishing time derivative
of the scalar field \cite{we,we1}. We see that in the case of piecewise exponential potentials such an effect is impossible, in spite of the presence of cusps. As a matter of fact the cusps, permitting such an exotic phenomenon should be
non-analytic \cite{we,we1}, while the cusps, which we treat now are still too smooth.

Concluding this section, we remark that choosing some simultaneous values of $\phi$ and $\dot{\phi}$, or, equivalently $\phi$ and $\phi'$ and
having the piecewise potential (\ref{piecewise}) one is able to reconstruct the whole past and future evolution of the
universe, using the general exact solutions, described in section 2. One can describe also the evolution of the phantom field
in the piecewise potential using the results of section 3. However, the effects of (de)-phantomization do not occur.
As was mentioned above, the effect of (de)-phantomization  requires the presence of strong non-analyticity of potential.
Besides, the field should approach the cusp with zero velocity and cross it. In the case considered in this section
instead of crossing we have a reverse motion.

\section{Conclusions}
As far as we know the only general exact cosmological solution in the presence of a scalar field, constructed explicitly, is the solution for
a flat Friedmann universe filled with a minimally coupled complex field with exponential potential.
While a particular exact solution for this case was known from 1985 \cite{power-law}, the general solution was constructed only in 1998 \cite{general}. In the present paper, we have studied accurately this solution, correcting some imprecisions, encountered
in the literature. In addition, we have generalized this solution for a piecewise exponential potential, which is continuous, but not smooth (with cusps). In spite of presence of cusps, in such a model  the effect of (de)-phantomization does not exist
\cite{we,we1} which requires non-analyticity of the potential.

\acknowledgments
A.K. was partially supported by the RFBR grant  11-02-643. A.A. was  supported by Grants RFBR 09-02-00073-a, 11-01-12103-ofi-m��and by SPbSU grant 11.0.64.2010 and also acknowledges the financial support from projects FPA2010-20807-C02-01 and 2009SGR502.

\end{document}